\documentclass[aps,prd,twocolumn,superscriptaddress]{revtex4-2}
\usepackage{CJK}

\usepackage{graphicx}
\usepackage{bm}
\usepackage{xcolor}
\usepackage{float}
\setcitestyle{numbers}
\usepackage[hidelinks]{hyperref}

\begin{document}

\preprint{APS/123-QED}

\title{Near-quantum-limited  axion dark matter search with the ORGAN experiment around 26 $\mu$eV}
\author{Aaron P. Quiskamp}
\email{aaron.quiskamp@uwa.edu.au}
\affiliation{Quantum Technologies and Dark Matter Laboratory, Department of Physics, University of Western Australia, 35 Stirling Highway, Crawley WA 6009, Australia.}
\author{Graeme R. Flower}
\affiliation{Quantum Technologies and Dark Matter Laboratory, Department of Physics, University of Western Australia, 35 Stirling Highway, Crawley WA 6009, Australia.}
\author{Steven Samuels}
\affiliation{Quantum Technologies and Dark Matter Laboratory, Department of Physics, University of Western Australia, 35 Stirling Highway, Crawley WA 6009, Australia.}
\author{Ben T. McAllister}
\email{ben.mcallister@uwa.edu.au}
\affiliation{Quantum Technologies and Dark Matter Laboratory, Department of Physics, University of Western Australia, 35 Stirling Highway, Crawley WA 6009, Australia.}
\affiliation{ARC Centre of Excellence for Dark Matter Particle Physics, Swinburne University of Technology, John St, Hawthorn VIC 3122, Australia}
\author{Paul Altin}
\affiliation{ARC Centre of Excellence For Engineered Quantum Systems, The Australian National University, Canberra ACT 2600 Australia}
\author{Eugene N. Ivanov}
\affiliation{Quantum Technologies and Dark Matter Laboratory, Department of Physics, University of Western Australia, 35 Stirling Highway, Crawley WA 6009, Australia.}
\author{Maxim Goryachev}
\affiliation{Quantum Technologies and Dark Matter Laboratory, Department of Physics, University of Western Australia, 35 Stirling Highway, Crawley WA 6009, Australia.}
\author{Michael E. Tobar}
\email{michael.tobar@uwa.edu.au}
\affiliation{Quantum Technologies and Dark Matter Laboratory, Department of Physics, University of Western Australia, 35 Stirling Highway, Crawley WA 6009, Australia.}

\date{\today}

\begin{abstract}
The latest result from the ORGAN experiment, an axion haloscope is presented. This iteration of the experiment operated at millikelvin temperatures using a flux-driven Josephson parametric amplifier for reduced noise, along with various other improvements over previous iterations. Covering the $25.45 - 26.27\,\mu\text{eV}$ ($6.15-6.35$ GHz) mass (frequency) range, this near-quantum limited phase of ORGAN employs a conducting rod resonator and a 7-T solenoidal magnet to place the most sensitive exclusion limits on axion-photon coupling in the range to date, with $|g_{a\gamma\gamma}| \gtrsim 2.8\times10^{-13}$ at a 95\% confidence level. 
\end{abstract}

\maketitle

\section{Introduction}
Dark matter remains one of the most elusive mysteries in the universe, accounting for approximately $85\%$ of its mass, yet it continues to evade detection \cite{planckcollaborationPlanck2018Results2020}. Among the candidates proposed to explain dark matter, axions emerge as a favoured solution due to their separate theoretical foundations and potential to resolve longstanding issues in particle physics. Originally introduced to address the strong \emph{CP} problem in quantum chromodynamics (QCD) \cite{pecceiCPConservationPresence1977, pecceiConstraintsImposedCP1977, wilczekProblemStrongInvariance1978, weinbergNewLightBoson1978}, axions bridge the worlds of cosmological phenomena and fundamental particle interactions. The Peccei-Quinn (PQ) symmetry breaking scale $f_a$ determines the axion's mass $m_a$, expected to be $\mathcal{O}(1\,\mu\text{eV}- 1\,\text{meV})$ \cite{turnerWindowsAxion1990, dineSimpleSolutionStrong1981, sikivieAxionCosmology2008} and coupling to Standard Model particles. The axion haloscope \cite{sikivieExperimentalTestsInvisible1983} is a method to detect axions in the laboratory by taking advantage of their expected coupling to photons. The haloscope method typically involves a cryogenic microwave cavity immersed in a strong static magnetic field (a source of virtual photons) to convert axions into real photons with a frequency $\nu_c$ related to the rest mass of the axion, $m_a \simeq h \nu_c /c^2$. The axion-converted photons will resonate inside the cavity if coupled to a geometrically appropriate resonant mode that satisfies the interaction Lagrangian: 

\begin{equation}
	\mathcal{L}_{a\gamma\gamma} = g_{a\gamma\gamma}a\vec{E}\cdot\vec{B},
	\label{L_int}
\end{equation}
 
\noindent where $a$ is the axion field, $\vec{E}$ is the electric field and $\vec{B}$ is the magnetic field (often externally applied). The coupling strength $g_{a\gamma\gamma}$ of this interaction is related to $f_a$ by 

\begin{equation}
	g_{a\gamma\gamma} = \frac{g_{\gamma}\alpha}{f_a \pi}.
\end{equation}

\noindent Here, $\alpha$ is the fine structure constant, and $g_\gamma$ is a dimensionless model-dependent parameter, taking a value of $-0.97$ and $0.36$ in the Kim-Shifman-Vainshtein-Zakharov (KSVZ) \cite{kimWeakInteractionSingletStrong1979, shifmanCanConfinementEnsure1980} and Dine-Fischler-Srednicki-Zhitnitsky (DFSZ) \cite{Zhitnitsky_1980tq, dineSimpleSolutionStrong1981} benchmark models, respectively. To date, only the ADMX \cite{duSearchInvisibleAxion2018b, braineExtendedSearchInvisible2020, bartramSearchInvisibleAxion2021a} and CAPP-12 TB \cite{yiAxionDarkMatter2023b, ahnExtensiveSearchAxion2024a} experiments have achieved sensitivity to DFSZ-coupled axions, partly due to the favourable sensitivity scaling at frequencies around $\sim 1\,\text{GHz}$. However, there is considerable interest in exploring the higher frequency parameter space (equivalently high axion mass), which is favoured when the PQ symmetry breaks after inflation \cite{borsanyiCalculationAxionMass2016, dineAxionsInstantonsLattice2017, buschmannDarkMatterAxion2022, ballesterosSeveralProblemsParticle2019}, with QCD lattice simulations indicating a lower bound for the axion mass (frequency) of $m_a\geq 14.6\,\mu\text{eV} \,(\nu_c\geq 3.54\,\text{GHz})$ \cite{berkowitzLatticeQCDInput2015}. The Oscillating Resonant Group AxioN (ORGAN) Experiment \cite{DirectSearchDarka, quiskampExclusionAxionlikeParticleCogenesis2024, mcallisterORGANExperimentAxion2017}, hosted in Western Australia, is among the growing list of higher-frequency haloscope experiments that attempt to search the post-inflationary mass window. Other notable experiments searching this range include HAYSTAC \cite{brubakerFirstResultsMicrowave2017, backesQuantumEnhancedSearch2021, jewellNewResultsHAYSTAC2023}, QUAX \cite{alesiniGalacticAxionsSearch2019, alesiniSearchInvisibleAxion2021}, MADMAX \cite{caldwellDielectricHaloscopesNew2017}, TASEH \cite{changTaiwanAxionSearch2022}, RADES \cite{melconAxionSearchesMicrowave2018a, alvarezmelconFirstResultsCASTRADES2021}, CAPP \cite{kim2024experimentalsearchinvisibledark} and GrAHal \cite{grenetGrenobleAxionHaloscope2021a}. \\

\begin{figure*}[t]
	\includegraphics[width=0.8\linewidth]{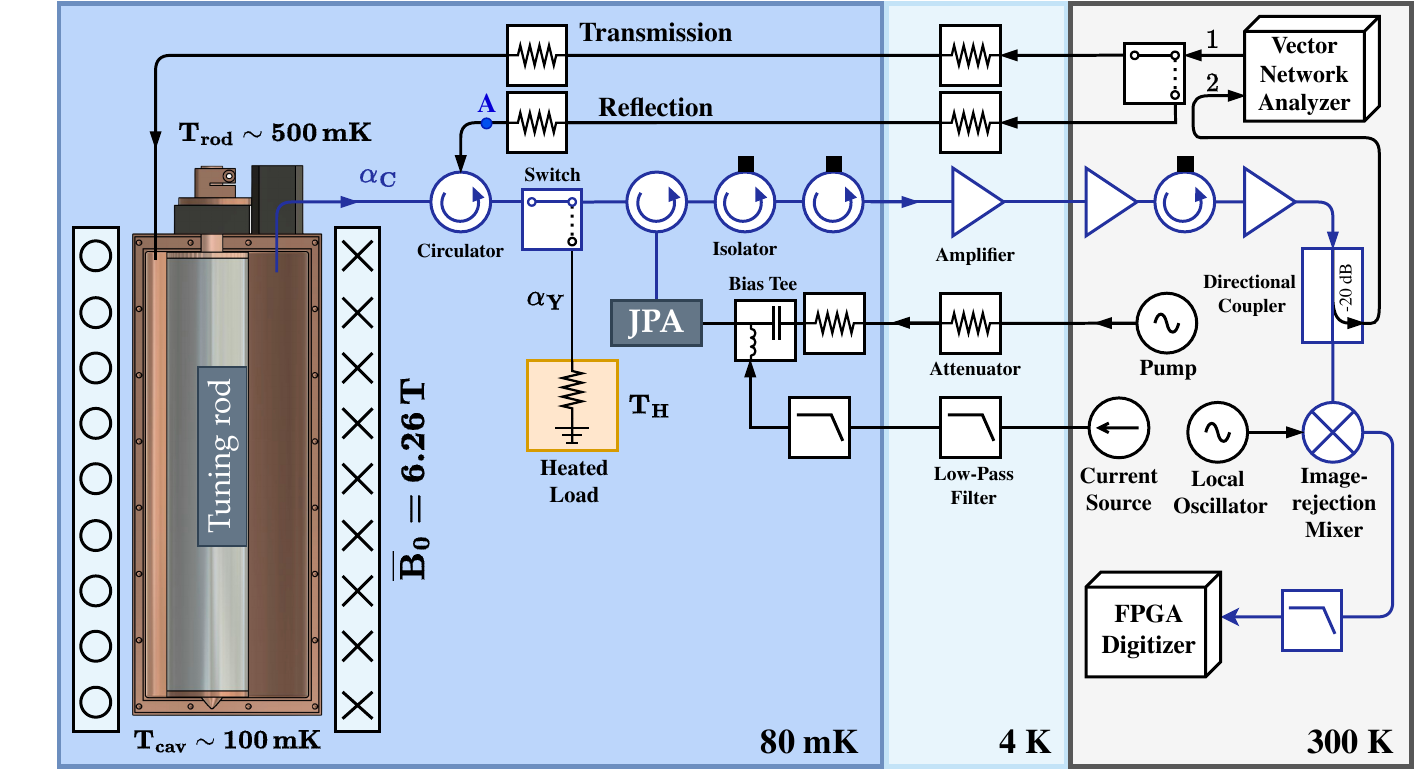}
	\caption{A diagram of the ORGAN-Q experiment, including the cavity, magnet, cryogenic receiver chain and room temperature data acquisition and calibration equipment. The coaxial cables and components shown in blue represent the path of a potential axion signal. The JPA is shown to be connected to the strongly coupled port of the cavity via a cryogenic RF switch and HEMT amplifier input using a circulator. The JPA is flux-pumped using a current source and driven by a pump tone. A vector network analyzer (VNA) is used to measure the frequency response of the $\mathrm{TM_{010}}$ mode in reflection and transmission via a room temperature RF switch. The signal from the cavity is down-mixed using an image rejection scheme for sampling by a field-programmable gate array (FPGA) digitizer. The linear and rotary piezomotors, shown in dark grey, control the antenna coupling and rod motion.}
	\label{OQ_diagram}
\end{figure*}

The signal power extracted from axion-photon conversion in a typical haloscope experiment can be expressed as \cite{brubakerFirstResultsMicrowave2017,changTaiwanAxionSearch2022}

\begin{equation}
	P_{\mathrm{sig}} = \left( g_{a\gamma\gamma}^{2} \frac{\rho_a}{m_a^2}\frac{\hbar^{3} c^{3}}{\mu_0} \right)\left(\omega_c B_{0}^{2} V C Q_L \frac{\beta}{1+\beta}  \right),
	\label{axion_power_OQ}
\end{equation}

\noindent where $\rho_a=0.45\,\mathrm{GeV/cm^3}$ is the local dark matter density \cite{Kafle_2014} (assumed to be all axions). The second set of parentheses denotes properties of the haloscope: $\omega_c=2\pi\nu_c$ is the angular resonant frequency, $B_0$ is the magnetic field strength, $V$ is the cavity volume, $C$ is the mode-dependent form factor that represents the coupling of the axion field to the electromagnetic mode, $Q_L$ is the loaded quality factor, and $\beta$ is the coupling strength of the strongly coupled antenna receiver. \\

In this work, we report on the most recent axion search with the ORGAN experiment, which scanned the $25.45 - 26.27\,\mu\text{eV}$ ($6.15-6.35$ GHz) mass (frequency) range of the axion-photon coupling parameter space. To achieve this, we implement a $\mathrm{TM_{010}}$-based conducting rod resonator that fully occupies our 7-T solenoidal magnet bore. The cavity has a volume of $0.708\, \text{L}$ and features a clamshell design made from oxygen-free high-conductivity copper. The aluminium tuning rod (selected as opposed to copper for its lower mass and therefore improved mechanical tuning), as shown in Fig. \ref{OQ_diagram}, results in a typical value for the form factor of $C\simeq0.42$ in the frequency range searched.\\

The ORGAN experiment is planned in several stages \cite{mcallisterORGANExperimentAxion2017} with the first phase of the experiment completed \cite{DirectSearchDarka, quiskampExclusionAxionlikeParticleCogenesis2024}. While this search falls outside ORGAN's targeted $15-50$ GHz frequency range, ORGAN Q serves as a test bed for a variety of techniques and technologies which may be applied in future ORGAN searches as Phase 2 of the experiment is prepared. These include a clamshell resonant cavity design, operating at millikelvin temperatures, employing a commercially available, flux-driven Josephson parametric amplifier (JPA) from Raytheon \cite{ranzaniWidebandJosephsonParametric2022b} before the HEMT amplifier (as opposed to solely HEMT amplifiers in Phase 1), the collection of high frequency resolution data (for a dedicated high resolution search), and the implementation of various other RF design features such as cryogenic switches and a heated load for noise calibration. The JPA consists of a nonlinear resonator, making use of Josephson junctions as its inductive element, and has a broadband impedance transformer on its input, which has been shown to increase the bandwidth beyond the gain bandwidth product \cite{ranzaniWidebandJosephsonParametric2022b, grebelFluxpumpedImpedanceengineeredBroadband2021, royBroadbandParametricAmplification2015}. It can be operated in either 3 or 4-wave mixing mode, has near-quantum limited performance and can produce broadband gain of order 100 MHz. In order to reduce the number of components and shift the pump tone away from the gain band, it is operated in 3-wave mixing mode, where the inductive element is parametrically modulated via a pump at roughly twice the signal frequency. Details of the JPA gain characterisation are discussed further in the Appendix A.\\

Implementing a near-quantum-limited amplifier at millikelvin temperatures greatly reduces the total system noise, $T_{\text{S}}$, compared to ORGAN Phase 1. For this reason, we refer to this spin-off phase of the experiment as ORGAN-Q, where Q stands for quantum. The total system noise, referred to that cavity output, is given by \cite{changTaiwanAxionSearch2022}:

\begin{equation}
	T_{\text{S}} = \tilde{T}_{\text{mx}} + (\tilde{T}_{\text{cav}} - \tilde{T}_{\text{mx}})L(\nu) + T_{\text{JPA}} + \frac{T_{\text{HEMT}}}{G_{\text{JPA}}},
\end{equation}

\noindent where $\tilde{T}_i = (\frac{1}{e^{h\nu/k_B T_i}-1} + \frac{1}{2})\times h\nu/k_B$ represents the sum of the effective temperature of the blackbody radiation at the physical temperature $T_i$, and the zero-point fluctuations of the vacuum, $T_{\text{cav}} \simeq 100\,\text{mK}$ is the physical temperature of the cavity, $L(\nu)$ is the cavity's Lorentzian lineshape, and $T_{\text{mx}} \simeq 80\,\text{mK}$ is the physical temperature of the attenuator $A$, anchored to the mixing chamber plate (as shown in Fig. \ref{OQ_diagram}). The effective noise added by the JPA, referred to the cavity output, is denoted as $T_{\text{JPA}}$. The gain $G_{\text{JPA}}$, of this initial amplification stage suppresses the noise added by the second-stage high electron mobility transistor (HEMT) amplifier, which is labelled as $T_{\text{HEMT}}$ and referred to the cavity output. \\

The sensitivity of an axion haloscope is given by the signal-to-noise ratio (SNR), which also determines the rate of scanning at a given axion-photon coupling exclusion power: 

\begin{equation}
	\text{SNR} = \frac{P_{\mathrm{sig}}}{k_B T_{\text{S}}}\sqrt{\frac{\tau}{\Delta\nu_a}}. 
\end{equation}

Here $\tau$ is the integration time, and $\Delta\nu_a \sim \nu_c/10^6$ is the expected axion linewidth due to the local axion kinetic energy distribution. For the standard halo model, this distribution is assumed to follow a Maxwell-Boltzmann profile \cite{turnerPeriodicSignaturesDetection1990a, brubakerHAYSTACAxionSearch2017}.\\

The total system noise is, therefore, a crucial parameter in determining the sensitivity of a haloscope. We calculate $T_{\text{S}}$ using the Y-factor method and the signal-to-noise ratio improvement (SNRI) method, which has become the standard calibration method in haloscope experiments \cite{bartramAxionDarkMatter2021, bartramSearchInvisibleAxion2021a, braineExtendedSearchInvisible2020, kwonFirstResultsAxion2021, yoonAxionHaloscopeUsing2022a}. The Y-factor method utilises the cryogenic switch, located on the mixing chamber plate as shown in Fig. \ref{OQ_diagram} to connect the receiver chain to a matched load, which can be heated at the ``$100\,\text{mK}$'' cold plate. The ``hot load'' equipped with a thermometer, is mounted to a piece of Nb and connected to the switch using a stainless steel coaxial cable to provide thermal isolation between the plates. By measuring the noise power, with the JPA pump tone off, at various hot-load temperatures $T_{\text{H}}$, we can determine $T^Y_{\text{HEMT}} = T_{\text{HEMT}} \frac{\alpha_C}{\alpha_Y}$, which is the measured noise temperature of the HEMT by Y-factor. To find the effective noise temperature contribution of the HEMT to signals at the output of the cavity, it is first scaled by the loss $\alpha_Y$ (Y-factor to switch) to refer this noise to the input of the switch, and then by $\alpha_C$ (cavity to switch) to refer it back to the cavity. These losses are shown in Fig. \ref{OQ_diagram} and were measured in a separate, dedicated cool down after the data-taking run was complete giving average values over the range of $\alpha_Y\,\sim\,-1.6$ dB and $\alpha_C\,\sim\,-1.5$ dB with some frequency dependence taken into account in all calculations. In order to simplify terminology, $T_{\text{HEMT}}$ includes the noise contribution from all losses between the switch and the HEMT input.\\

\begin{table}[t]
\begin{tabular}{ll}
\hline\hline
Axion sensitive data taking\hspace{5em} & $\sim 17$ days                                       \\
Frequency range                         & $6.15-6.35$ GHz                                       \\
Axion mass $m_a$                        & $25.45-26.27$ $\mu$eV                                 \\
Frequency bin width                     & 954 Hz                                              \\
Tuning step                             & $\sim \Delta\nu_c/5$                                 \\
Sweep time $\tau$                       & $\sim 30-60$ minutes                                       \\
Magnetic field (average)                & $6.26 \pm 0.01$ T                                              \\
Q factor (typical)                      & 1700 ($\sigma\,=\,400$)                                 \\
Form factor (typical)                   & $0.42\pm 0.02$                                           \\
Cavity volume                           & 0.708 L                                             \\
Cavity coupling $\beta$ (typical)       & 1.45 ($\sigma\,=\,0.32$)                                \\
MXC temperature                         & $\sim 80$ mK                                               \\
Cavity temperature                      & $\sim 100$ mK                                              \\
Rod temperature                         & $\sim 500$ mK                                        \\
Effective HEMT noise (typical)          & 9.3 K ($\sigma\,=\,0.87$ K)                                             \\
System noise (typical)                  & 928 mK ($\sigma\,=\,108$ mK)                            \\ 
System noise quanta @ 6.2 GHz           & 3.12 ($\sigma\,=\,0.36$)                         \\  \hline\hline                    
\end{tabular}
\caption{Parameters for the experimental run}
\label{param_table}
\end{table}

When attempting to measure $T_{\text{JPA}}$ using the Y-factor method, saturation of gain was observed even at low $T_{\text{H}}$ values due to the limited dynamic range and broad gain-bandwidth of the JPA. Thus, we instead determine $T_{\text{S}}$ (which includes $T_{\text{JPA}}$) using the SNRI method. This method involves a spectrum comparison that must be done relatively quickly for each tuning-rod (cavity frequency) position by comparing the SNR with the JPA on (pump tone on) to the SNR with the JPA off (pump tone off). The SNRI is given by \cite{bartramAxionDarkMatter2021}

\begin{equation}
	\text{SNRI} = \left(\mathrm{\frac{S_{on}}{N_{on}}}\right)\left(\mathrm{\frac{S_{off}}{N_{off}}}\right)^{-1},
\end{equation}

\noindent where $\mathrm{S_{on}}$ ($\mathrm{S_{off}}$) are the transfer functions measured using a vector network analyzer (VNA) with the JPA turned on (off), with $\mathrm{G_{JPA}} = \mathrm{S_{on}}/\mathrm{S_{off}}$ and $\mathrm{N_{on}}$ ($\mathrm{N_{off}}$) the noise powers measured using a spectrum analyzer with the JPA turned on (off). The JPA transfer function given by the S parameter $S_{21}$, is measured by operating a room temperature switch that switches the VNA output (port 1) to the RF ``reflection'' line to measure the frequency response at the VNA input (port 2). Using the measured SNRI and $T^Y_{\text{HEMT}}$, the total system noise temperature referred to the cavity output is given by \cite{ahnExtensiveSearchAxion2024a}

\begin{equation}
	T_{\text{S}} = \frac{T^Y_{\text{HEMT}}\frac{\alpha_Y}{\alpha_C} + \tilde{T}_{\text{mx}}}{\text{SNRI}},
\end{equation}

\noindent where all parameters are measured quantities. A summary of relevant experimental parameters is given in Table \ref{param_table}.\\

\section{Results}

\begin{figure}[t]
		\includegraphics[width=\linewidth]{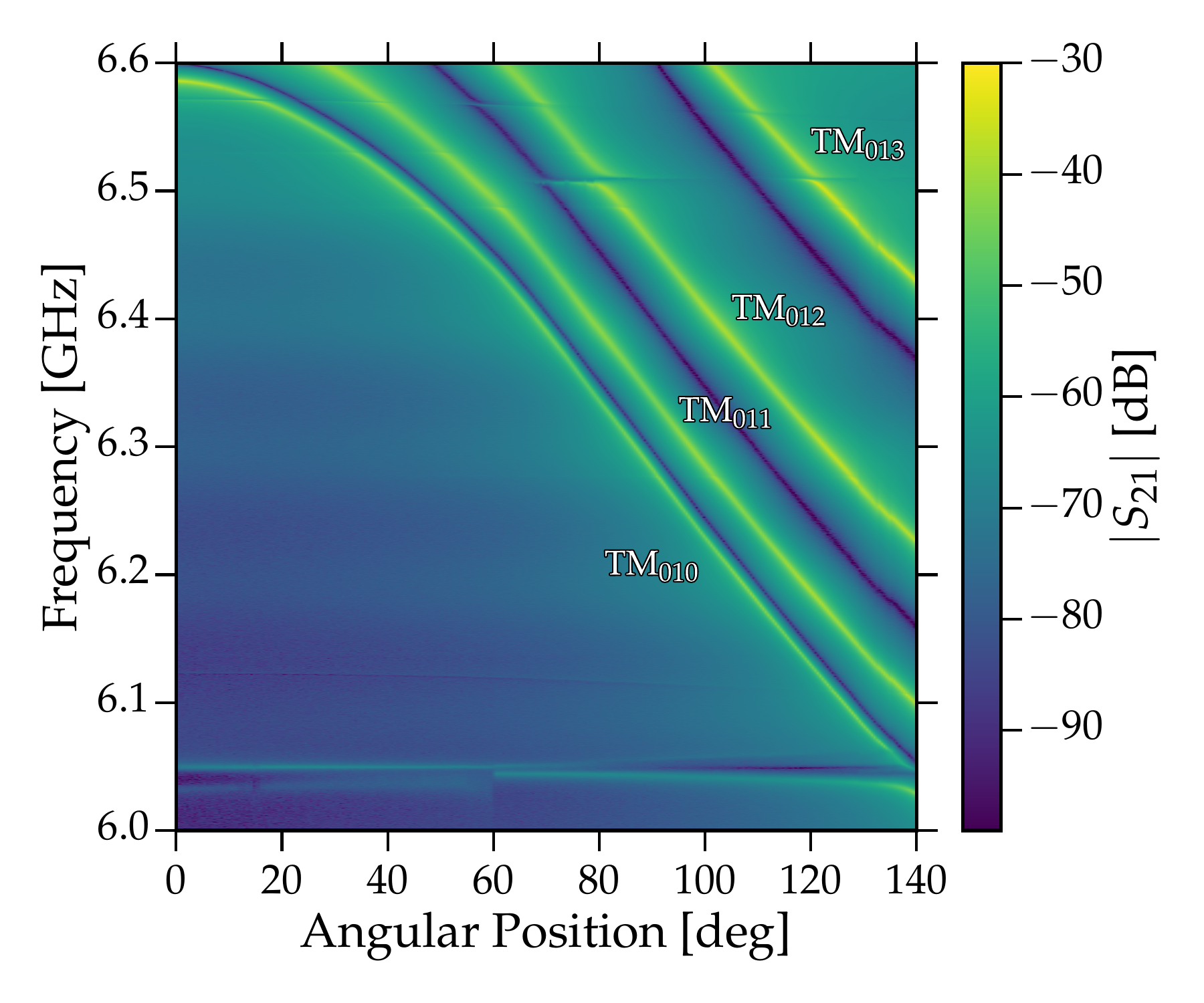}
		\caption{A colour density plot of the transmission coefficient $|S_{21}|$ (dB) as a function of resonant frequency (GHz) and angular position (degrees) of the tuning rod. The $\mathrm{TM_{010}}$ mode is annotated and shown to tune between $6.15-6.35$ GHz without any mode-crossings, which appear as horizontal lines. The higher-order $\mathrm{TM_{01p}}$ modes tune in the same way and are also annotated. Lighter regions indicate greater transmission, while darker regions indicate lower transmission.}
		\label{OQ_mode_map}
\end{figure}

We acquired axion-sensitive data for $\sim 17$ days, to search for axions in a region without mode crossings, with masses between $25.45 - 26.27\,\mu\text{eV}$ ($6.15-6.35$ GHz). A total of 489 resonant frequencies were scanned with an integration time $\tau\sim \,30-60\,\text{minutes}$. Rotating the off-axis tuning rod via an Attocube ANR240 piezoelectric stepper motor shifted the resonant frequency in step sizes of $\sim\Delta\nu_c/5$, as shown in Fig. \ref{OQ_mode_map}. The tuning rod was hollowed out in such a way that the centre of mass coincided with the axis of rotation, minimizing the torque required to move the rod, which ensured smooth tuning at cryogenic temperatures. The stepper motor was directly coupled to the tuning rod axle, while the bottom rod end cap was positioned in a conical groove (as shown in Fig. \ref{OQ_diagram}) to avoid radiation leakage and provide additional thermalisation between the rod and cavity. The antenna coupling was adjusted using an Attocube ANPz101 piezoelectric stepper motor mounted to the cavity, which is also shown in Fig. \ref{OQ_diagram}. \\

At each frequency, the parameters $Q_L$, $\beta$ and SNRI are measured by operating the room temperature switch to take measurements of the VNA transfer function. Throughout the scan, $Q_L$ had a typical value of $\sim 1700$ $(\sigma=400)$, and $\beta$ was adjusted to maintain above critical coupling, with an average value of $\sim 1.45 \,(\sigma=0.32)$. Quality factors likely limited by radiation leakage due to the tuning rod, something we intend to address in future experiments. \\

As mentioned, $T_{\text{S}}$ is determined at each cavity frequency and varied as a function of detuning from resonance. We express the system noise temperature in the $j$th bin of the $i$th spectrum as $T^{\text{S}}_{ij}$, with an average value of $\sim 928\,\text{mK}\,(\sigma=108 \,\text{mK})$ over the scanned region. This corresponds to a total system noise quantum of 3.12 ($\sigma=0.36$) at 6.2GHz. This is much higher than the noise temperature expected assuming $T_{\text{JPA}}\sim 380\,\text{mK}$, as measured by Raytheon \cite{ranzaniWidebandJosephsonParametric2022b}. Unfortunately, tuning the rod position with the rotary piezomotor led to significant heating of the rod, resulting in a typical rod temperature of $T_{\text{rod}} \sim 500 \,\text{mK}$. We attribute the higher than expected value for $T_{\text{S}}$ to the ``hot rod'', which is accounted for in the SNRI measurements. \\

Data acquisition is fully automated using Python and LabVIEW. The appropriate JPA gain curve is chosen based on the cavity centre frequency $\nu_c$, maintaining a gain greater than 18 dB over most of the searched frequency range. Once amplified, the thermal noise (and potential axion signal) is down-mixed to an intermediate frequency (IF) using a local oscillator and an IQ mixer. The outputs of the IQ mixer are combined using a $90^{\circ}$ hybrid coupler to achieve image rejection of the amplifier-only sideband. The signal is then sampled by a 250 MS/s digitizer (NI-5761R), and a zero dead-time fast Fourier transform is processed on a field-programmable gate array (FPGA; Xilinx Kintex-7, NI-7935R).\\ 

\begin{figure*}[t]
		\includegraphics[width=0.8\linewidth]{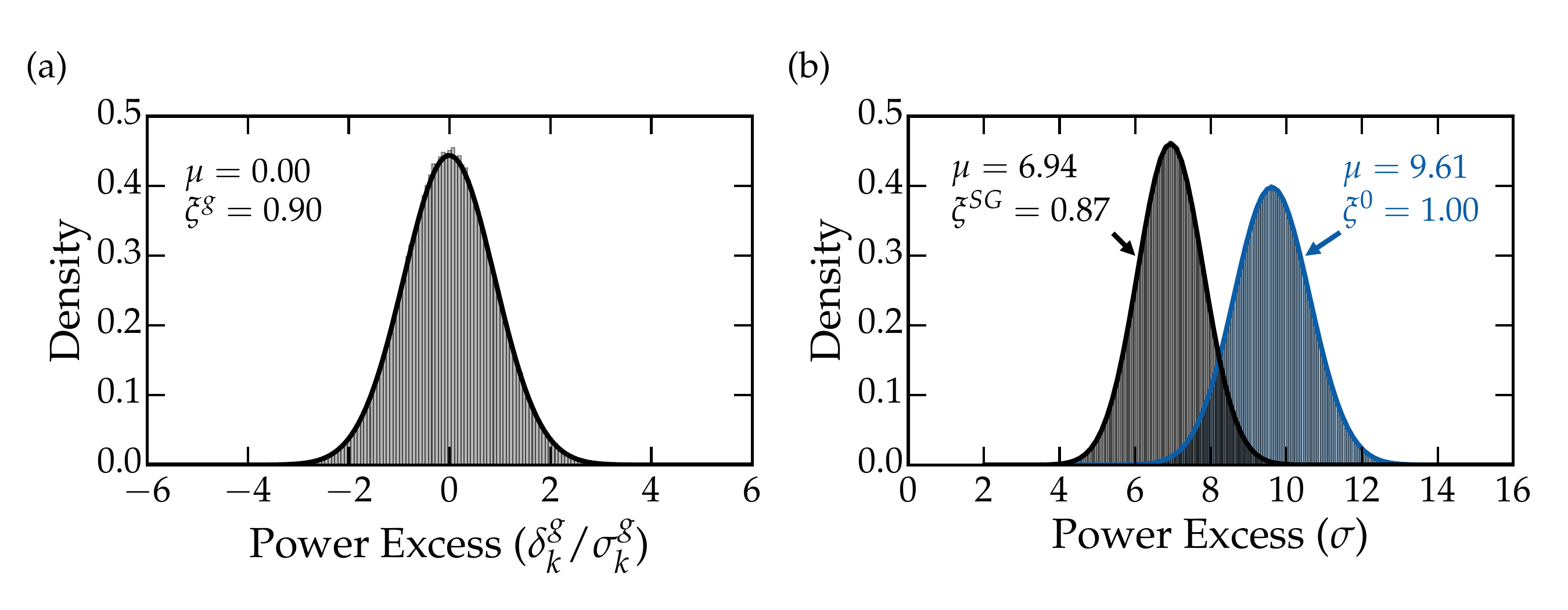}
		\caption{(a) The distribution of all grand spectrum bins $\delta^g_k/\sigma^g_k$, after being SG filtered, vertically combined with overlapping spectra and finally horizontally combined according to the expected Maxwell-Boltzmann lineshape. The distribution deviates from the expected normal Gaussian distribution, with a reduced width of $\xi^g=0.90$, which is attributed to the negative correlations induced by the SG filter. (b) Histograms of the SNR for a synthetic axion signal injected into Gaussian white noise, with one dataset SG-filtered (black) to simulate the real data analysis pipeline and the other dataset unfiltered (blue). The SG-filtered distribution is narrower in a similar way to the real data in panel (a), with $\xi^{SG}=0.87$. The SG-filter-induced attenuation of an axion signal is given by the ratio of the two means $\eta_{\text{SG}}=6.94/9.61=0.72$. }
		\label{hist_OQ}
\end{figure*}

In contrast to previous ORGAN searches \cite{DirectSearchDarka, quiskampExclusionAxionlikeParticleCogenesis2024}, ORGAN-Q operates at a low enough frequency to capture ``high-resolution'' spectra, which are sensitive to nonvirialized axion flows. These narrow, cold axion flows have been proposed in the caustic ring model \cite{sikivieCausticRingsDark1998, sikivieEvidenceRingCaustics2003, duffyCausticRingModel2008, chakrabartyImplicationsTriangularFeatures2021} or as a consequence of the QCD axion being born after inflation \cite{ohareAxionMiniclusterStreams2023}. To achieve this, we operate two parallel data streams on the FPGA. The first stream, similar to previous ORGAN searches \cite{DirectSearchDarka, quiskampExclusionAxionlikeParticleCogenesis2024}, uses real-time FPGA-based averaging to generate a 6554 point, 6.25-MHz-wide spectrum centred at 45 MHz, resulting in a bin width of $\Delta\nu_b\approx 954\,\text{Hz}$. The second stream does not perform any averaging and operates at one-quarter of the sampling frequency as the main data stream, producing a 3,355,444 point, 1.5625-MHz-wide spectrum, with a fractional spectral width of $\mathcal{O}(10^{-10})$. The results of the high-resolution analysis will be presented in a forthcoming publication.\\

As in previous ORGAN searches \cite{DirectSearchDarka, quiskampExclusionAxionlikeParticleCogenesis2024}, we follow the HAYSTAC analysis procedure \cite{brubakerHAYSTACAxionSearch2017,yiAnalyticalEstimationSignal2023}. The analysis window is restricted to an IF bandwidth of $\sim 3\,\text{MHz}$, and the cropped spectra are subsequently normalized using a Savitsky-Golay (SG) filter with a 101-point window and a polynomial degree of 4. The SG filter removes the slowly varying baseline from the raw spectra while preserving narrow spectral features on the scale of the axion width $\Delta\nu_a\sim 6.25\,\text{kHz}$. Overlapping power spectra are then vertically combined using a maximum likelihood weighted sum of contributing spectra to maximize the SNR. The weights for this sum are determined according to the expected axion power in the $j$th IF bin of the $i$th spectrum and the bin-dependent $T^{\text{S}}_{ij}$ value. \\

The ``grand power spectrum'' is constructed by optimally filtering sets of nine consecutive bins with the expected Maxwellian lineshape in the boosted lab frame, retaining $\sim90\%$ of the axion signal \cite{brubakerHAYSTACAxionSearch2017}. The standard deviation for the $k$th grand spectrum bin is then the quadrature weighted sum of contributing combined spectra standard deviations, denoted as $\sigma^g_k$. The grand power spectra are normalized to $\sigma^g_k$ and shown histogrammed in panel (a) of Fig. \ref{hist_OQ}. The width of this distribution is reduced to $\xi^g=0.90$ due to the negative correlations induced by the SG filter between neighboring bins. The SG filter is also known to attenuate potential axion signals and is a function of the choice of filter parameters. This effect is simulated 1,000,000 times by injecting synthetic axions into the ORGAN-Q analysis pipeline and comparing it with unfiltered data \cite{brubakerHAYSTACAxionSearch2017, DirectSearchDarka}. The result is shown in panel (b) of Fig. \ref{hist_OQ}, where the ratio of the two means $\eta_{\text{SG}}=6.94/9.61=0.72$ gives the SG-filter-induced attenuation. This factor is reduced by $\xi^g$ to give $\eta_{\text{SNR}}=\eta_{\text{SG}}/\xi^g = 0.80$, which represents the total reduction in the SNR of each grand bin to a given axion signal \cite{brubakerHAYSTACAxionSearch2017}. To validate the simulation, we also compute $\eta_{\text{SG}}$ analytically based on recent work in \cite{yiAnalyticalEstimationSignal2023}, and find $\eta_{\text{SG}}=0.73$.\\ 

\begin{figure*}
		\includegraphics[width=0.8\linewidth]{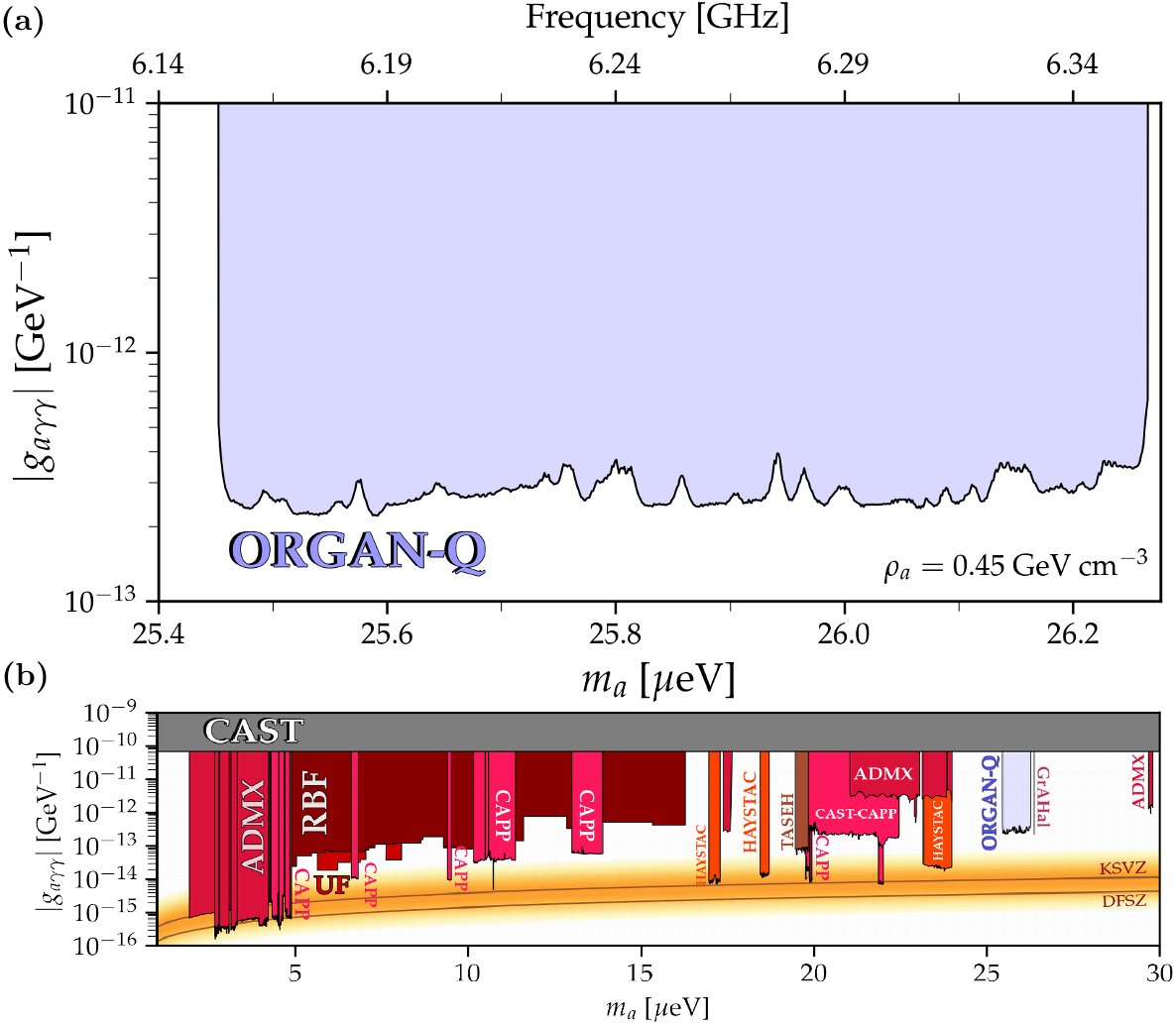}
		\caption{(a) Axion-photon coupling exclusion limits at a 95\% confidence level for ORGAN-Q (light purple). (b) Compares these results with existing limits from other experiments \cite{castcollaborationNewCASTLimit2017, bartramSearchInvisibleAxion2021a, braineExtendedSearchInvisible2020, duSearchInvisibleAxion2018b, bartramDarkMatterAxion2023, boutanPiezoelectricallyTunedMultimode2018, brubakerFirstResultsMicrowave2017, yiAxionDarkMatter2023b, kwonFirstResultsAxion2021, kimNearQuantumNoiseAxionDark2023a, changTaiwanAxionSearch2022, grenetGrenobleAxionHaloscope2021a} collated using resources from C. O'Hare's GitHub \cite{AxionLimits}. The ORGAN-Q limits are the most sensitive in this region. The KSVZ \cite{kimWeakInteractionSingletStrong1979, shifmanCanConfinementEnsure1980} and DFSZ \cite{Zhitnitsky_1980tq, dineSimpleSolutionStrong1981} axion models are also shown.}
		\label{OQ_limits}
\end{figure*}

We chose to forgo rescans due to multiple magnet quenches during the data-taking run, which significantly extended the experimental time. Bin-by-bin exclusion limits on axion-photon coupling are placed based on the maximum normalized power excess of $6.68\sigma$, which corresponds to an SNR target of $8.33\sigma$ at 95\% confidence level \cite{brubakerHAYSTACAxionSearch2017}. The resulting limits for the $25.45 - 26.27\,\mu\text{eV}$ mass range are the most sensitive to date and are shown in Fig. \ref{OQ_limits}, assuming axions make up 100\% of the local dark matter density of $0.45 \,\mathrm{GeV/cm^3}$. The fractional uncertainty on the axion-photon coupling  $\delta g_{a\gamma\gamma}$ is calculated similarly to previous analyses \cite{DirectSearchDarka, quiskampExclusionAxionlikeParticleCogenesis2024}. We find $\delta g_{a\gamma\gamma} \approx 6.3\%$, which is dominated by uncertainties in the parameters $T_{\text{S}}\,(11.6\%)$ and $C\,(5\%)$. Further details on the experimental uncertainties can be found in the Appendix B.

\section{Conclusion}
We have reported the results from the first near-quantum-limited phase of the ORGAN experiment, operated at millikelvin temperatures and with a commercially available JPA. It is the most sensitive probe on the axion-photon coupling strength in the mass range between $25.45 - 26.27\,\mu\text{eV}$, excluding $|g_{a\gamma\gamma}| \gtrsim 2.8\times10^{-13}$ at a 95\% confidence level. \\

\begin{acknowledgments}
This work is supported by Australian Research Council Grants CE17010009 and CE200100008 as well as the Australian Government’s Research Training Program.
\end{acknowledgments}

\section*{Appendix A: JPA characterisation}
\begin{figure*}[t!]

    \includegraphics[width=0.8\linewidth]{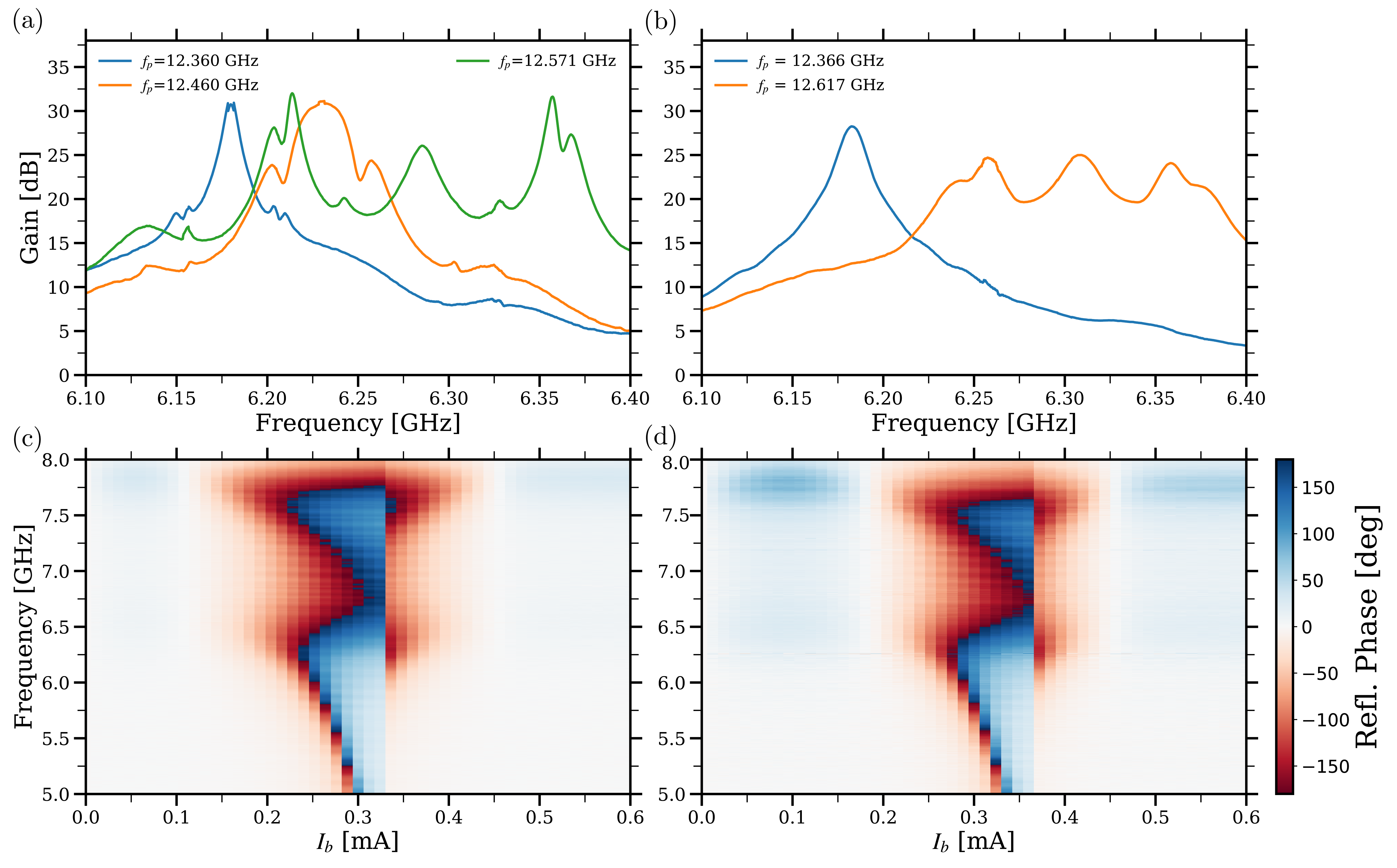}
	\caption{JPA characterisation. a),c) JPA gain for the matching parameters in Table \ref{JPAtable} and normalised reflected phase showing resonance tuning as a function of DC current bias and VNA frequency respectively for 12th Feb - 27th Feb. b),d) The same for 6th March - 10th March.}
	\label{jpa_char}
\end{figure*}

Further details as to the characterisation and use of the JPA are provided. As discussed in the above, the first stage amplifier used in the experimental run was a commercially available JPA from Raytheon \cite{ranzaniWidebandJosephsonParametric2022b}. Due to the nonstandard nature of the input coupling  (a broadband impedance transformer to the junction resonator), the JPA  is operated in a different way to typical capacitively coupled JPAs. This has the significant benefit of producing broadband gain beyond the gain bandwidth product -- but does complicate optimisation, as best gain and noise performance is often achieved with pump tones significantly detuned from resonance. \\

\begin{table}[h]
\centering
\setlength{\tabcolsep}{11pt}
\begin{tabular}{llll}
\hline\hline
$I_b$ (mA) & $f_b$(GHz) & $P_p$ (dBm) & $f_{\text{range}}$ (GHz) \\
\hline
\multicolumn{4}{l}{12th Feb -- 27th Feb}                             \\
0.235      & 12.36       & -48.9       & 6.15 -- 6.195                 \\
0.228      & 12.46       & -41.6       & 6.195 -- 6.275                \\
0.228      & 12.571      & -39.7       & 6.275 -- 6.35                 \\
\hline
\multicolumn{4}{l}{6th March -- 10th March}                          \\
0.264      & 12.366      & -49.1       & 6.14 -- 6.20                  \\
0.252      & 12.617      & -40.6       & 6.2 -- 6.35                  \\
\hline\hline
\end{tabular}
\label{JPAtable}
\caption{JPA tuning parameters prior to and following the magnet quench on the 27th of February 2024.}
\end{table}

The junction is in a SQUID configuration, with flux through the loop driven by the pump port, while DC current applied to the flux port tunes the JPA resonance. Due to the resonance being highly overcoupled to a 50 $\Omega$ load, it is characterized using the normalised reflected phase as shown in Fig. \ref{jpa_char}. \\

As the device is cooled from above the transition temperature of aluminium (approximately 1.2 K) to below, it traps a random quantity of flux in the SQUID loop resulting in small changes to the tuning parameters of the device compared to previous cools, requiring recharacterisation each time the device makes that transition. Unfortunately, due to magnet quenches that occurred during the run resulting in a rapid warm up of the dilution refrigerator, the JPA required recharacterisation. A list of JPA pump and bias parameters used over particular frequencies is presented in Table \ref{JPAtable}. Additionally, gains associated with these pump and bias parameters are noted in Fig. \ref{jpa_char}.

\section*{Appendix B: Uncertainty Analysis}
The systematic uncertainties relevant to the experimental run are summarised in Table I of the main text. The uncertainty in the frequency-dependent form factor $C$ is assumed to be within the range reported for similar experiments, typically $3\%$ to $5\%$ \cite{bartramAxionDarkMatter2021, bartramSearchInvisibleAxion2021a}, from which we adopt the upper limit of $5\%$. The uncertainty in the frequency-dependent total system noise temperature $T_{\text{S}}$ is taken to be the standard deviation of its measured values. This approach captures the variability in $T_{\text{S}}$ across the frequency range of interest, reflecting the fluctuations in the measured SNRI and $T\mathrm{^Y_{HEMT}}$. We find the uncertainty in $T_{\text{S}}$ to be $\sim 11.6\%$. The parameters $Q_L$, $\beta$, and $B_0$ were measured with high precision and contribute a negligible uncertainty, $\leq 1\%$

Considering only the dominant sources of uncertainty, the fractional uncertainty in the axion-photon coupling, $\delta g_{a\gamma\gamma}$ is given by

\begin{equation}
    \frac{\delta g_{a \gamma \gamma}}{g_{a \gamma \gamma \gamma}} \approx \sqrt{\left(\frac{1}{2} \frac{\delta T_{\mathrm{S}}}{T_{\mathrm{S}}}\right)^2+\left(\frac{1}{2} \frac{\delta C}{C}\right)^2} \approx 6.3 \%
\end{equation}

\bibliographystyle{apsrev4-1}

\end{document}